\documentclass[11pt]{article}
\pdfoutput =1
\usepackage{graphics}
\usepackage{graphicx}
\usepackage{subcaption}
\DeclareGraphicsExtensions{.pdf}
\usepackage{float} 
\usepackage{subfloat}
\usepackage{amsmath}
\usepackage{slashed}
\usepackage{amsfonts}  
\usepackage{amssymb}
\usepackage{physics}

\def\th{\theta}

\usepackage{comment}
\usepackage{color}

\usepackage{subcaption}

\begin{document}
\date{}
\title{{\bf{\Large Probing sine dilaton gravity with flow central charge}}}
\author{
{\bf { ~Paramesh Mahapatra}$
$\thanks{E-mail:  paramesh\_m@ph.iitr.ac.in}}\\
 {\normalsize  Department of Physics, Indian Institute of Technology Roorkee,}\\
  {\normalsize Roorkee 247667, Uttarakhand, India}
\\[0.3cm]
{\bf { Hemant Rathi}$ $\thanks{E-mail: hemant.rathi@saha.ac.in}}\\ 
{\normalsize Saha Institute of Nuclear Physics, 1/AF Bidhannagar, }\\ 
{\normalsize Kolkata 700064, India }\\
{\normalsize School of Basic and Applied Sciences, K.R. Mangalam University,}\\
{\normalsize Gurugram 122103, Haryana, India}
  \\[0.3cm]
 {\bf { ~Dibakar Roychowdhury}$
$\thanks{E-mail:  dibakar.roychowdhury@ph.iitr.ac.in}}\\
 {\normalsize  Department of Physics, Indian Institute of Technology Roorkee,}\\
  {\normalsize Roorkee 247667, Uttarakhand, India}
\\[0.3cm]
}
\maketitle
\abstract{We construct a holographic c-function for sine dilaton gravity (sDG) in the domain wall gauge. We show the equivalence between sDG and the two copies of Liouville conformal field theory (LCFT) and compute the associated central charge. We reproduce the central charge of LCFT from the holographic c-function of sDG corresponding to the UV limit. In contrast, in the deep IR limit, the c-function flows to the pure JT gravity, where the central charge becomes identically zero}.  

\section{Overview and motivation}
The concept of a c-function was first introduced by Zamolodchikov in 1986 in the context of two-dimensional (2D) field theories\footnote{For generalisation to even-dimensional field theories, see \cite{Cardy:1988cwa}.} \cite{Zamolodchikov:1986gt}. It has been demonstrated that for 2D field theories, a c-function exists along the Renormalization Group (RG) flow. This function yields the central charge \cite{Kiritsis:2019npv}, which measures the degrees of freedom of the associated Conformal Field Theories (CFT) at their respective fixed points. In other words, the c-function serves as a measure that tracks these degrees of freedom throughout the RG flow. Following this, several important studies were conducted on the c-function in the context of RG flow \cite{Cappelli:1990yc}-\cite{Anselmi:1997rd}.

Notably, the c-function has been explored in the context of holography by the authors in  \cite{Alvarez:1998wr}. The natural candidate for a holographic c-function is a monotonically decreasing function (from UV to IR) of the radial coordinate $r$, which serves as a measure of the energy scale of the boundary CFT in a generic gauge/gravity duality \cite{Natsuume:2014sfa}.


 
 In recent years, several methods have been proposed for constructing the holographic c-function  \cite{Myers:2010tj}-\cite{Deddo:2023pid}. One approach involves using holographic entanglement entropy, as demonstrated by the authors in \cite{Myers:2012ed}-\cite{Albash:2011nq}. Another method considers the renormalization group (RG) flow within a domain-wall background and employs a superpotential to construct the c-function, as outlined in \cite{Suh:2020qnl}-\cite{Freedman:1999gp}. Furthermore, the authors in \cite{Alkac:2018whk}-\cite{Deddo:2023pid} motivate a parallel construction of the c-function based on the null energy condition (NEC).

In this paper, we aim to understand the holographic c-function within the context of sine-dilaton gravity (sDG) \cite{Blommaert:2024ydx}-\cite{Blommaert:2024whf} in two-dimensional space-time, which is holographically dual to the double-scaled Sachdev-Ye-Kitaev (DSSYK) model  \cite{Blommaert:2023opb}-\cite{Berkooz:2024lgq}. The sDG theory is a Einstein gravity in two dimensions coupled with a sine dilaton potential. For more works on sDG, refer to \cite{Blommaert:2025avl}-\cite{Blommaert:2023wad}. On the other hand, DSSYK is a one-dimensional quantum mechanical framework of interacting Majorana fermions, where both the number of fermions ($N$) and the number of interactions ($p$) tend to infinity while keeping the ratio $p^2/N$ fixed. 

It is worth noting that, in the low-energy limit\footnote{See Section 5, for a detailed discussion.}, the sDG theory reduces to the well-known Jackiw-Teitelboim (JT) gravity  \cite{Jackiw:1984je}-\cite{Teitelboim:1983ux}. The boundary description for the JT gravity is provided by the Sachdev-Ye-Kitaev (SYK) \cite{Sachdev:1992fk}-\cite{Maldacena:2016hyu} model\footnote{On the field theory counterpart, the (holographic) RG flow corresponds to a flow from a $q$-Schwarzian theory (or DSSYK) in the UV to the ordinary SYK in the IR. For a detailed discussion, see section 4.1.}. In this paper, we construct the holographic c-function that computes the central charge of the sDG theory in the ultraviolet (UV) limit of a holographic RG flow. This is achieved in two steps. As step one, we map sDG to two copies of Liouville CFTs and compute its central charge. Next, we show that this central charge can be obtained from a systematic construction of a c-function and identifying its appropriate UV limit. The computation of the c-function has been carried out following the methods of \cite{Suh:2020qnl} and \cite{Alkac:2018whk}, which are shown to be mutually compatible.

The rest of the paper is organized as follows.

\begin{itemize}
              
        \item In Section 2, we discuss the sDG theory and obtain the solutions for background fields in the static gauge. We then present these solutions in the Domain Wall or Conformal gauge \cite{Suh:2020qnl} using an appropriate coordinate transformation.
        
        \item In Section 3,  we provide a brief overview of the Liouville Conformal Field Theory (LCFT). Next, we map the sDG theory into two copies of LCFTs and then compute the central charges for both sectors of the LCFT.
        
        \item In Section 4, we construct the c-function associated with the LCFTs using the superpotential approach \cite{Suh:2020qnl} and the null energy condition \cite{Alkac:2018whk}. We also demonstrate their mutual compatibility.
        
        \item In Section 5, we show that the sDG theory reduces to JT gravity in the deep IR limit. Finally, we summarize our findings and conclude our work by outlining some interesting future directions.
         \end{itemize}

\section{Sine dilaton gravity in domain wall gauge}\label{rev}
In this Section, we provide a brief review of sine dilaton gravity (sDG) and obtain the solutions for background fields \cite{Blommaert:2024ydx}-\cite{Blommaert:2024whf}. We further demonstrate that the space-time metric of the sDG can be mapped to a conformally AdS space-time \cite{Blommaert:2024whf}. Finally, we express these solutions in the domain wall gauge \cite{Suh:2020qnl} using a suitable coordinate transformation. This will be useful for computing the c-function \cite{Suh:2020qnl} associated with sDG in the following sections.

The action for sDG in two space-time dimensions can be expressed as \cite{Blommaert:2024ydx}-\cite{Blommaert:2024whf}
\begin{equation}
    S_{sDG} = \frac{1}{2}\int d^2x\sqrt{-g}\left( \Phi R + \frac{\sin(2|\log q| \Phi)}{|\log q|}\right) +\int d\tau \sqrt{h}\left( \Phi K - i\frac{e^{-i|\log q| \Phi}}{2|\log q|} \right), \label{sdgact0}
\end{equation}
where $\Phi$ is the dilaton and $R$ is the Ricci scalar in two dimensions. The last two terms are the Gibbons Hawking York boundary term and the counter term respectively. Here $K$ is the trace of extrinsic curvature. The action is supplemented by the following boundary conditions \cite{Blommaert:2024ydx}-\cite{Blommaert:2024whf}
\begin{equation}
    \sqrt{h} e^{i\Phi_{bdy}/2} = i\hspace{1mm},\hspace{2mm} \Phi_{bdy} = \frac{\pi}{2} + i \infty. \label{bdy}
\end{equation}

Notice that, $|\log q|$ is a parameter that has its origin in the holographic description of the sDG model, known as the double-scaled SYK (DSSYK) model \cite{Blommaert:2023opb}-\cite{Berkooz:2024lgq}. This parameter is also related to the central charge of the corresponding Liouville conformal field theory (LCFT) \cite{Verlinde:2024zrh}.

Without loss of generality, we now rescale the dilaton as $2|\log q| \Phi \rightarrow \Phi$, yielding the following action
\begin{equation}
    S_{sDG} = \frac{1}{4|\log q|}\int d^2 x \sqrt{-g}\left( R \Phi + 2 \sin\Phi\right)  + \frac{1}{2|\log q|}\int \sqrt{h}\left( \Phi K - i e^{-i\Phi/2} \right).\label{sdgact}
\end{equation}

Next, we vary the above action (\ref{sdgact}) with respect to both the space-time metric ($g_{\mu\nu}$) and the dilaton ($\Phi$), resulting in the following set of equations of motion
\begin{align}
    (\bigtriangledown_\mu\bigtriangledown_\nu - g_{\mu\nu}\square)\Phi + g_{\mu\nu}\sin\Phi = 0,\label{eom1}\\
    R + 2\cos\Phi = 0.\label{eom2}
\end{align}

Notice that the above set of equations (\ref{eom1})-(\ref{eom2}) can be easily solved in a gauge where the space-time can be expressed as \cite{Blommaert:2024ydx}-\cite{Blommaert:2024whf}
\begin{align}
    ds^2 =- F(r)dt^2 + \frac{dr^2}{F(r)}.\label{le1}
\end{align}

In the above gauge (\ref{le1}),  the equations (\ref{eom1})-(\ref{eom2}) simplify to 
\begin{align}
F''(r) - 2 \cos\Phi(r) =&\hspace{1mm} 0\label{r1e},\\
 \Phi''(r) =&\hspace{1mm} 0. \label{metric2}
\end{align}

The solutions to the above equations (\ref{r1e})-(\ref{metric2}) are given below
\begin{align}
    F(r) =&\hspace{1mm} 2\cos \theta - 2 \cos r,\label{metricsol}\\
    \Phi(r) =&\hspace{1mm} r,\label{phisolr}
\end{align}
where the boundary of the radial contour of integration is at $ r = \pi/2 + i \infty$, and the holographic screen is located at $r = \pi/2$ \cite{Blommaert:2024ydx}-\cite{Blommaert:2024whf}. Here, $\theta$ is the location of the black hole horizon.

Notice that in the $r$ coordinates, the boundary of the space-time is located at an imaginary point. We can transform the space-time metric (\ref{le1}) into a conformally $ AdS_2 $ metric by applying the following coordinate transformation \cite{Blommaert:2024whf}
\begin{equation}
    r = \frac{\pi}{2} + i \log(\rho + i \cos \theta).\label{ct}
\end{equation}

Substituting (\ref{ct}) into (\ref{le1}), we obtain
\begin{align}
    ds^2=e^{i\Phi(\rho)}\left[-\left(\rho^2-\rho_h^2\right)dt^2+\frac{d\rho^2}{\left(\rho^2-\rho_h^2\right)}\right]
    =e^{i\Phi(\rho)}ds^2_{AdS_2},\label{cads2}
\end{align}
where $\rho_h=\sin\th$ is the location of the horizon and the boundary of space-time is located at $\rho\rightarrow\infty$.\par
The complexification of the radial coordinate ($r$) therefore maps the sDG metric to a conformally $AdS_2$ space-time (\ref{cads2}) and also makes the dilaton complex valued. The $AdS_2$ structure of the bulk geometry is crucial to establish the holographic dictionary between the sDG and the DSSYK. The DSSYK correlators exhibit similar structure to that of a boundary to boundary propagator of a massive particle in $AdS_2$, indicating that the bulk dual should also have an $AdS_2$ structure. It has been shown in \cite{Blommaert:2024whf} that a non-minimally coupled massive scalar that is characterized by the following action

\begin{align}
    S_{probe} = \int d^2x \sqrt{-g} \left(g^{\mu\nu} \partial_\mu \phi \partial_\nu \phi + m^2e^{-i\Phi} \phi^2 \right),
\end{align}
is an appropriate probe to reproduce the DSSYK correlator. The complex valued coupling $e^{-i \Phi}$ (with the complex valued dilaton) together with the complex sDG metric (\ref{cads2}) produce an ``effective" $AdS_2$ geometry experienced by the probe scalar, whose action is given below
\begin{align}
    S_{probe} =  \int d^2x \sqrt{-g_{eff}} \left(g^{\mu\nu}_{eff} \partial_\mu \phi \partial_\nu \phi + m^2\phi^2 \right),
\end{align}
where $g_{eff}^{\mu\nu}$ is the metric of $AdS_2$. In other words, the probe experiences real $AdS_2$ geodesic motion which in turn produces real propagators that match with the DSSYK correlators \cite{Blommaert:2024whf}.

In the present analysis, we are interested in computing the holographic central charge, the flow equations, and the c-function associated with the sDG model (\ref{sdgact}). This analysis can effectively be conducted in the domain wall gauge \cite{Suh:2020qnl}. Therefore, our next aim is to map the space-time metric (\ref{cads2}) into the domain wall background. \par 
 To achieve this, we employ the following coordinate transformation
\begin{align}
     u = \frac{1}{2\rho_h}\log\left( \frac{\rho+ \rho_h}{\rho-\rho_h}\right).\label{ct1}
\end{align}
Notice that in the $u$ coordinate, the boundary of the space-time is located at $u\rightarrow0$.

After substituting (\ref{ct1}) into (\ref{cads2}), we obtain the following domain wall background 
\begin{align}\label{dwbg}
    ds^2=e^{2A(u)}(-dt^2 + du^2),
\end{align}
where\footnote{One can show that (\ref{auct}) and (\ref{auct1}) solve the equations of motion (\ref{eom1})-(\ref{eom2}) in the domain wall gauge (\ref{dwbg}) whose solutions are given in (\ref{auct})-(\ref{auct1}).} 
\begin{align}
    A(u) =&\hspace{1mm} \frac{1}{2}\log\left(\frac{\rho_h^2}{\sinh^2(\rho_h u)}\right) + \frac{i}{2}\Phi(u),\label{auct}\\
     \Phi(u)=&\hspace{1mm} \frac{\pi}{2} + i \log\left(\rho_h \coth(\rho_h u) +  \sqrt{ \rho_h^2-1} \right). \label{auct1}
\end{align}
For the rest of our analysis, we work in the domain wall background described by the equations (\ref{dwbg})-(\ref{auct1}).

\section{The Liouville CFT and the central charge}
In the following, we show the equivalence between sDG and the Liouville CFT (LCFT) \cite{DiFrancesco:1997nk} and compute the central charge of sDG (\ref{sdgact}) using this analogy. We begin by a quick review of the LCFT.

\subsection{Review of LCFT}\label{rlcft}

The Liouville CFT or the Liouville gravity emerges in the context of non-critical string theories \cite{Polyakov:1981rd}. In these theories, the dynamics of the space-time metric ($g_{\mu\nu}$) is effectively described by a conformal factor known as the dilaton ($\Phi$), which appears in the following line element
\begin{align}
    ds^2 = g_{\mu\nu}dx^\mu dx^\nu=e^{2\Phi}\hat{g}_{\mu\nu}dx^\mu dx^\nu, \label{ds}
\end{align}
where $\hat{g}_{\mu\nu}$ is the flat space-time metric. This is similar in spirit to the domain wall metric (\ref{dwbg}).

The action for an LCFT in two space-time dimensions \cite{DiFrancesco:1997nk} is given by
\begin{align}
    S_{LCFT} = \frac{1}{4\pi}\int d^2x \sqrt{-g} \left( (\partial\Phi)^2 + QR\Phi + 4\pi\mu e^{2b\Phi}\right) +\frac{1}{2\pi}\int dx \sqrt{-h} (QK\Phi + 2\pi \mu_b e^{b\Phi}). \label{Laction}
\end{align}
Here, $R$ is the Ricci scalar in two dimensions, $b$ is the Liouville coupling, $\mu$ is the cosmological constant, $\mu_b$ is the boundary cosmological constant, and $Q = b+b^{-1}$ is the background charge. The background charge is required for the cancellation of the quantum trace anomaly \cite{DiFrancesco:1997nk}. Given a 2D gravity theory that allows a solution in the form (\ref{ds}), it could be recast as (\ref{Laction}) using appropriate field redefinitions. We show this explicitly in the case of sDG.

The stress-energy tensor for a LCFT \cite{DiFrancesco:1997nk} can be expressed as
\begin{align}\label{stl}
    T(z) = - :(\partial \Phi)^2:(z) + Q\partial^2 \Phi(z).
\end{align}

Next, we compute the operator product expansion (OPE) \cite{DiFrancesco:1997nk} of the stress-energy tensor (\ref{stl}), which yields the following
\begin{align}\label{stlope}
    T(z)T(\omega) = [-(\partial \Phi)^2(z)][-(\partial \Phi)^2 (\omega)] + Q^2\partial^2\Phi(z)\partial^2 \Phi(\omega) + \text{mixed terms}.
\end{align}
The first term in the above expression (\ref{stlope})  represents the contribution coming from the free scalar part of the action ($S_L$)(\ref{Laction}), while the second term accounts for the correction due to the background charge term.

One can further simplify the expression given in (\ref{stlope}) by using the definition of the operator product expansion for a scalar field \cite{DiFrancesco:1997nk}
\begin{align}
    \Phi(z)\Phi(\omega) \approx -\frac{1}{2}\log|z-\omega|^2. \label{sope}
\end{align}

Using (\ref{sope}) into (\ref{stlope}), one finds
\begin{align}
    T(z)T(\omega)\approx \frac{c/2}{(z-\omega)^4} + \frac{2T(\omega)}{(z-\omega)^2} + \frac{\partial T(\omega)}{(z-\omega)} + ...,\label{ctformula}
\end{align}
where $c$ is the central charge associated with the LCFT (\ref{Laction}), and is given by \cite{DiFrancesco:1997nk}
\begin{align}
    c = 1 + 6Q^2. \label{cdefn}
\end{align}




\subsection{Mapping sDG to LCFT}\label{maps}
In this Section, we map the sDG theory (\ref{sdgact}) into two copies of the LCFTs (\ref{Laction}). Next, we compute the central charge for the sDG theory (\ref{sdgact}) by comparing it with the central charge of the LCFT (\ref{ctformula}).

In order to map the sDG theory (\ref{sdgact}) to the Liouville CFT (\ref{Laction}), we consider the following conformal gauge (\ref{dwbg})
\begin{align}\label{confmet}
    g_{ab} = e^{2A} \hat{g}_{ab},
\end{align}
where $(a,b)$ are the 2D indices and $\hat{g}_{ab}$ is a flat space-time metric. 

 Substituting (\ref{confmet}) into (\ref{sdgact}), we obtain
\begin{align}
    S_{sDG} = \frac{1}{4|\log q|}\int d^2 x \left(2 \partial \Phi \partial A + 2 e^{2A} \sin \Phi\right) -\frac{i}{4|\log q|} \int d\tau ( e^{A-i\Phi/2}).\label{ma}
\end{align}

Next, we propose the complex field redefinition as follows \cite{Blommaert:2024ydx}
\begin{align}
     A = &\hspace{1mm}\frac{Z_1 + iZ_2}{4}, \label{defZ1} \\
     \Phi = &\hspace{1mm}\frac{Z_1 - iZ_2}{2i}. \label{defZ2}
\end{align}

Substituting (\ref{defZ1})-(\ref{defZ2}) into (\ref{ma}), we find\footnote{Notice that the Ricci Scalar ($\hat{R}$) vanishes in the flat gauge $\hat{g}_{ab} = \eta_{ab}$.}
\begin{align}
    S_{sDG} = &\hspace{1mm}\int d^2x \left(\frac{1}{4\pi} \left((\partial Z_1)^2 + 4\pi\mu_1 e^{2b_1Z_1} \right) + \frac{1}{4\pi} \left((\partial Z_2^2) + 4\pi \mu_2 e^{2b_2 Z_2} \right)\right) \nonumber\\      &+ \int \textcolor{blue} {d\tau (\mu_{b} e^{b_2 Z_2})}\nonumber\\
    =&\hspace{1mm} S_r(Z_1) + S_l(Z_2) \label{lact1}
\end{align}
where we make the following identifications\footnote{Here, we have rescaled the fields $Z_1\rightarrow Z_1/\sqrt{\frac{\pi}{4|\log q|i}}$ and $Z_2\rightarrow Z_2/\sqrt{\frac{\pi}{4|\log q|i}}$ to match with the canonical form of the Liouville CFT (\ref{Laction}).}
\begin{align}
    b_1=\sqrt{\frac{|\log q| i}{\pi}}\hspace{1mm},\hspace{2mm}b_2=i\sqrt{\frac{|\log q|i}{\pi}} \hspace{1mm},\hspace{2mm}\mu_1=\frac{1}{4|\log q| i}\hspace{1mm},\hspace{2mm}\mu_2=-\frac{1}{4|\log q| i}\hspace{1mm},\hspace{2mm} \mu_{b} = \frac{-i}{2|\log q|}.\label{id}
\end{align} 

 Here, $S_r$ and $S_l$ are respectively the right and left sectors of the Liouville CFT. The total central charge of the sDG theory is the sum of the central charges of both sectors of the LCFTs.
 
 It is important to notice that the first boundary condition (\ref{bdy}) can be expressed as $e^{b_1 Z_1} = i$, which is a constant quantity. As a result, the field $Z_1$ experiences a fixed length boundary condition, therefore the associated boundary cosmological term in the action (\ref{lact1}) vanishes identically \cite{Blommaert:2024ydx}, \cite{Mertens:2020hbs}. Notice that the equivalence between the Liouville CFTs and sDG that is shown here, holds at the \emph{classical} level. However, a similar equivalence between LCFT and sin-hyperbolic dilaton gravity has been studied beyond the classical level by the authors in \cite{Mertens:2020hbs}-\cite{Fan:2021bwt}. These papers can be used to serve as a basis while extending the present analysis beyond classical regime and in particular to explore various issues like the partition function, the path integral measures and the anomaly subtleties. For example, one can examine the action (\ref{lact1}) using the first order formalism, which reveals the underlying quantum group theoretic structure. Subsequently, the gravitational matrix representation can be utilized to compute the amplitudes. Finally, these quantities can be compared with the corresponding amplitudes and partition function in sine-dilaton gravity \cite{Mertens:2020hbs}.

 We next compute the central charge for the right sector ($S_r$) of the LCFT. Using the identifications (\ref{id}), the square of background charge for the right sector ($Q_1^2$) (\ref{Laction}) can be expressed as
\begin{align}
    Q_1^2 = (b_1+b_1^{-1})^2=2 + \frac{|\log q|i}{\pi} - \frac{\pi i}{ |\log q|}. \label{Qs}
\end{align}

After substituting (\ref{Qs}) into (\ref{cdefn}), the central charge ($c^{(r)}$) for the right sector of the LCFT can be expressed as 
\begin{align}
    c^{(r)} = 13 + 6i \left( \frac{|\log q| }{\pi} - \frac{\pi}{|\log q|}\right). \label{rcdef}
\end{align}

A similar calculation shows that the central charge for the left sector of the LCFT is complex conjugate of (\ref{rcdef}) and is given by
\begin{align}
    \bar{c}^{(l)} = 13 - 6i \left( \frac{|\log q| }{\pi} - \frac{\pi}{|\log q|}\right). \label{lcdef}
\end{align}

 The total central charge ($c$) is the sum of (\ref{rcdef}) and (\ref{lcdef}) and is given by   
\begin{align}
    c = c^{(r)} + \bar{c}^{(l)} = 26.\label{cls}
\end{align}
The above result (\ref{cls}) exactly matches with the central charge of sDG obtained by the authors in \cite{Blommaert:2024ydx}. 






\section{c-function for sDG}
In this Section, we construct the holographic c-function associated with the RG flow from sDG in the UV (expressed in the form (\ref{lact1})) to JT gravity in the IR. In particular, we adopt two methods to construct the c-function. The first method involves using the superpotential ($W$) approach of \cite{Suh:2020qnl}, while the second method is based on the Null Energy Condition (NEC) approach of \cite{Alkac:2018whk}. We emphasize that the construction of our proposed c-function is a bottom-up approach. We do not present any derivation. Rather, we apply the methods used in \cite{Suh:2020qnl} and \cite{Alkac:2018whk} to the construction. These methods are based on well-established physical foundations and have been explored in higher dimensions \cite{deBoer:1999tgo}-\cite{Freedman:1999gp}. Finally, we demonstrate the mutual compatibility between these two approaches, and in particular show that the central charge (\ref{cls}) follows from the appropriate definition of the c-function in the UV of a holographic RG flow.


\subsection{The superpotential approach}\label{sw}

In the superpotential approach \cite{Suh:2020qnl}, we write down the equations of motion associated with the gravitational action $S(g_{\mu\nu},\Phi,\phi)$  in the domain wall gauge (\ref{dwbg}). Here, $g_{\mu\nu}$ is the space-time metric, $\Phi$ is the dilaton field, and $\phi$ is a scalar field. Notably, these equations are second-order differential equations. However, one can convert them into first-order differential equations by employing a function known as the superpotential ($W$) \cite{Suh:2020qnl}.  \par
Typically, these first order equations could be schematically expressed as
\begin{gather}
    \Phi' e^{-A} = f(W, \partial_\Phi W ), \label{deq}\\
   \phi'e^{-A} =g(W,\partial_\Phi W),\\
    \frac{d A}{du} e^{-A} = h(W, \partial_\Phi W), \label{weq}
\end{gather}
where $f,g$ and $h$ are functions of superpotential ($W$) and its derivative ($\partial_{\Phi}W$).
\par In the literature, these first-order equations (\ref{deq})-(\ref{weq}) are referred as flow equations. By using the superpotential ($W$), we can finally define a quantity that decreases monotonically, known as the c-function \cite{Suh:2020qnl}\footnote{The authors in \cite{Suh:2020qnl} compute the monotonic decreasing function or the c-function particularly for the 2D dilaton gravity.}. In the following, we compute the c-function associated with sDG (\ref{lact1}), following the above methodology.


To begin with, we vary the action (\ref{lact1}) and obtain the following equations of motion  
\begin{align}
     Z_1'' - 2e^{Z_1}=0,\label{ee1}\\
    Z_2'' + 2i e^{iZ_2}=0,\label{ee2}
\end{align}
Here $'$ denotes the derivative with respect to the variable $u$. 

Next, we convert the second-order differential equations (\ref{ee1})-(\ref{ee2}) into a set of first-order flow equations 
\begin{align}
     Z_1' = 2e^{(Z_1+ iZ_2)/4} \mathcal{W}(Z_1,Z_2),\label{e1d1}\\
     Z_2' = -2ie^{(Z_1+iZ_2)/4}  \bar{\mathcal{W}}(Z_1,Z_2),\label{e1d2}
\end{align}
where we propose the following superpotential
    \begin{align}
        \mathcal{W}(Z_1,Z_2)= e^{(Z_1 - iZ_2)/4}.\label{W}
    \end{align}

To verify the consistency of our results, we substitute the complex superpotential ($\mathcal{W}(Z_1,Z_2)$) (\ref{W}) into the flow equation (\ref{e1d1})-(\ref{e1d2}). This leads to the following equations
\begin{gather}
    Z_1' = 2e^{Z_1/2},\label{e1d3}\\
    Z_2' = -2ie^{iZ_2/2}.\label{e1d4}
\end{gather}
Now it is straightforward to check that the derivatives of the above equations (\ref{e1d3})-(\ref{e1d4}) yield the equations of motion (\ref{ee1})-(\ref{ee2}).

Furthermore, one can obtain the solutions for $Z_1$ and $Z_2$ by solving the equations (\ref{e1d3}) and (\ref{e1d4}), which results in the following
\begin{align}
    Z_1(u) = \log(c_1 \text{csch}^2(c_1 u)) + i\pi - 2\log\left(c_1 \coth(c_1 u) + i \sqrt{1-c_1^2}\right),\label{zsol1}\\
    Z_2(u) =  -i\log(c_1 \text{csch}^2(c_1 u)) - \pi + 2i\log\left(c_1 \coth(c_1 u) - i \sqrt{1-c_1^2}\right)\label{zsol2},
\end{align}
where $c_1$ is an integration constant. After comparing equations (\ref{zsol1})-(\ref{zsol2})  (by means of the definitions (\ref{defZ1})-(\ref{defZ2})) with the equations (\ref{auct})-(\ref{auct1}), one can identify $c_1 = \rho_h$. 

Next, we define a monotonic density function $D(u)$. Using the superpotential (\ref{W}), this turns out to be\footnote{Here, we would like to emphasize that the monotonicity of the c-function follows strictly by taking the modulus of $D(u)$. This is our proposal to construct the holographic c-function for 2D dilaton gravity, which has been constructed along the lines of \cite{Suh:2020qnl}. Consequently, this definition should be considered a result of a bottom-up approach.}
\begin{align} \label{ed}
    D(u) = |4 e^{3(Z_1+iZ_2)/4} \mathcal{W}^2|= |e^{(Z_1 + iZ_2)/4} Z_1'^2|=\frac{4\rho_h^3\text{csch}^3(\rho_h u)}{(1 + \rho_h^2 \text{csch}^2(\rho_h u))^{3/2}},
\end{align}
where we substitute the expression for $Z_1$ and $Z_2$ using (\ref{zsol1})-(\ref{zsol2}). 

 Notice that our definition of the density function ($D(u)$) (\ref{ed}) closely parallels to that of \cite{Suh:2020qnl}, where the derivative of c-function is defined using the flow equations and a superpotential. In higher dimensions, the flow equations typically arise from the  Hamilton-Jacobi equations of the gravitational theories and can be directly related to the Callan-Symanzik equations of the boundary QFT in the asymptotic limit \cite{deBoer:1999tgo}. The authors in \cite{Suh:2020qnl} explored this concept in the context of 2D gravitational theory, analogous to the higher-dimensional theories. They expressed the c-function as a composition of the kinetic contributions of the dynamical fields in the theory, with an overall exponential prefactor. Following similar line of arguments as in \cite{Suh:2020qnl}, we construct $D(u)$ using the kinetic contribution of the dynamical field ($Z_1$) and an exponential prefactor as illustrated in (\ref{ed}). Below, we use this $D(u)$ to construct a function (\ref{cfw}) whose derivative is negative (decreasing from UV to IR (\ref{cdm})) and thereby satisfying all the necessary criteria of a holographic c-function.


Using (\ref{ed}), the c-function for the right and left sectors of LCFT (\ref{lact1}) can be expressed as
\begin{align}
    c(u) =  \frac{c^{(r)}}{N}\int_u^\infty D(u) du\hspace{1mm},\hspace{2mm} \bar{c}(u) =  \frac{\bar{c}^{(l)}}{N} \int_u^\infty D(u) du,\label{cm}
\end{align}
where $c^{(r)}$ and $\bar{c}^{(l)}$ are the central charges as defined in (\ref{rcdef}) and (\ref{lcdef}) respectively. In what follows, we show that these are the central charges in the UV of a holographic RG flow, which is identified as sDG. \par Furthermore, here $N$ is a positive normalization constant given by
\begin{align}
    N = \int_0^\infty D(u)du. \label{Norms}
\end{align}

As previously outlined, the sDG theory can be equivalently expressed as two sectors of LCFTs (\ref{lact1}) using the complex field redefinition (\ref{defZ1})-(\ref{defZ2}). The complex nature of the coordinate transformation leads to complex central charges (\ref{rcdef})-(\ref{lcdef}) for each sector \cite{Blommaert:2024ydx}. As a result, the corresponding c-functions $c(u)$ and $\bar{c}(u)$ (\ref{cm}) also emerge as complex quantities. However, it is important not to interpret the individual sector c-functions as independent physical flow central charges \cite{Blommaert:2024ydx}. Instead, they characterize the RG flow of the two complexified Liouville sectors (\ref{lact1}). The real and physically meaningful holographic c-function associated with the RG flow can be expressed as\footnote{Notice that the integral (\ref{cfw}) is a real quantity. Furthermore, we discuss the asymptotic behavior of this integral in Appendix \ref{app}.}
\begin{align}\label{cfw}
    \hat{c}(u) = c(u) + \bar{c}(u) = \frac{c}{N}\int_u^\infty\frac{4\rho_h^3 \text{csch}^{3}(\rho_h v) dv}{(1 +\rho_h^2\text{csch}^2(\rho_h v))^{3/2}}.
\end{align}
where $c$ is the total central charge (\ref{cls}) in the UV.


It is important to note that the functions $c(u)$ and $\bar{c}(u)$ (\ref{cm}) are defined in such a way that, in the UV limit, the integrals in the numerator exactly equal the normalisation constant ($N$ (\ref{Norms})). This yields the central charges for both sectors, namely, $c^{(r)}$ (\ref{rcdef}) and $\bar{c}^{(l)}$ (\ref{lcdef}), respectively. In other words, near the UV scale\footnote{See Appendix \ref{app} for detailed expansion. } in the holographic RG flow, the c-function (\ref{cfw}) yields the total central charge (\ref{cls}) of sDG.

\begin{figure}[h!]
    \centering
    \begin{subfigure}[b]{0.4\textwidth}
       \includegraphics[height = 0.30\textheight, width = 1.5\textwidth]{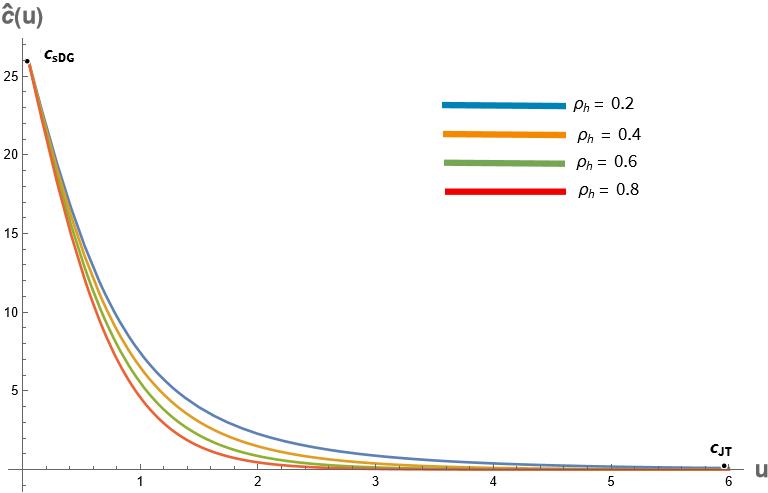}
        \caption{Plot of $\hat{c}(u)$ against $u$ for different values of $\rho_h$.}
        \label{fig1a}
    \end{subfigure}
    \hfill
    \begin{subfigure}[b]{0.4\textwidth}
         \includegraphics[height = 0.34\textheight, width = 1\textwidth]{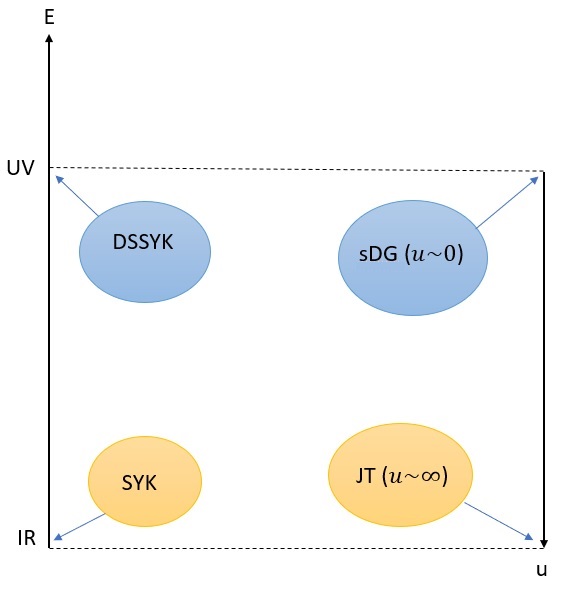}
        \caption{Holographic RG flow.}
        \label{fig1b}
    \end{subfigure}
    \caption{In Figure (1a), $c_{sDG} = c^{(r)} + \bar{c}^{(l)}$ corresponds to the central charge of sDG in the UV of the holographic RG flow. On the other hand, $c_{JT}$ is the central charge in JT gravity. Figure (1b) shows the RG flow from UV to IR and its holographic counterpart in the bulk.}
    
\end{figure}

To analyze the monotonicity of the c-function (\ref{cfw}), we take the derivative of the c-function with respect to the variable $u$, which yields
\begin{align}\label{cdm}
 \frac{d}{du} [\hat{c}(u)] = -\frac{4\rho_h^3 \text{csch}^{3}(\rho_h u)}{(1 +\rho_h^2\text{csch}^2(\rho_h u))^{3/2}}\left( \frac{c}{N} \right).
 \end{align}
 Notice that the derivative (\ref{cdm}) of the c-function is negative, which reveals the monotonicity as shown in Figure (\ref{fig1a}).

To clarify this behaviour further, we plot (\ref{cfw}) against the radial coordinate $u$ for various choices of the parameter $\rho_h$ as shown in Figure (\ref{fig1a}). Notice that the c-function (\ref{cfw}) approaches the UV central charge (i.e. $c=26$) (\ref{cls}) (or equivalently the central charge of sDG) in the limit $u\rightarrow0$, while it vanishes in the deep IR regime $u\rightarrow\infty$. We argue that the IR regime corresponds to pure JT gravity, which arises as the low-energy limit of the sDG theory \cite{Blommaert:2024ydx}-\cite{Blommaert:2024whf}, \cite{Mahapatra:2025fpx}.

Clearly, we have an interpolating geometry which interpolates between the sDG in the UV and the ordinary JT gravity in the IR. In the dual picture, this corresponds to a flow from DSSYK in the UV to SYK in the IR (see Figure (\ref{fig1b})).

Before proceeding further, we discuss the notion of central charge in JT gravity. Notice that, in 2D CFT, the central charge is typically calculated using the operator product expansion (OPE) of the stress-energy tensor \cite{Kiritsis:2019npv}, \cite{Rathi:2024qsy}. The concept of central charge arises from the central extension of the Witt algebra to the Virasoro algebra \cite{Kiritsis:2019npv}, \cite{Rathi:2024qsy}. To determine the central charge in two dimensions, we analyze the variational properties of the stress-energy tensor under infinitesimal coordinate transformations and apply the operator product expansion \cite{Kiritsis:2019npv}, \cite{Rathi:2024qsy}. The one-dimensional result can be obtained by setting one variable to zero. In this framework, one-dimensional CFT can be viewed as the chiral half of a two-dimensional CFT defined on a strip, where only one copy of the Virasoro algebra remains \cite{Hartman:2008dq}.

Alternatively, one can compute the associated central charge holographically by investigating the variational properties of the 1D boundary stress-energy tensor ($T_{zz}(z)$) under an infinitesimal coordinate transformation \cite{Mahapatra:2025fpx}, \cite{Castro:2008ms}-\cite{Rathi:2021aaw}. After performing all the intermediate calculations, the variation of $T_{zz}(z)$ can be expressed as 
\begin{equation}
    \delta_\epsilon T_{zz}(z) = \frac{c}{12}\partial_z^3\epsilon(z) + 2\partial_z\epsilon(z)T_{zz}(z) + \epsilon(z)\partial_zT_{zz}(z),\label{apt}
\end{equation}
where $\epsilon(z)$ is the infinitesimal parameter and $c$ is the central charge associated with the 1D CFT.


\par
In the following subsection, we compute the c-function using the NEC approach \cite{Alkac:2018whk} and demonstrate that it equals with the c-function as obtained using the superpotential method of \cite{Suh:2020qnl}.
\subsection{The NEC approach}

It is interesting to notice that one can also construct the c-function using the Null Energy Condition (NEC) as demonstrated by the authors in \cite{Alkac:2018whk}. The NEC can be used to define the monotonicity of the c-function, as has been done in \cite{Alkac:2018whk}. The condition is given by
\begin{equation}
    T_{\mu\nu}\xi^\mu\xi^\nu\ge 0
\end{equation}
where $\xi^\mu$ is an arbitrary null-vector and $T_{\mu\nu}$ is the stress-energy tensor \cite{Penrose:1964wq}. 
\par By choosing an appropriate null-vector, the NEC in the domain wall gauge\footnote{ See Appendix B for a derivation of the physical NEC without gauge fixing.} can be written as
\begin{equation}
    T^t_t - T^r_r \le 0
\end{equation}
in the Lorentzian signature. 
\par For a generic gravitational theory in $n$-dimensions with the action
\begin{equation}
    S = \int d^nx \sqrt{-g} \left(  L_{gr} - \frac{1}{2}\partial_\mu \phi \partial^\mu \phi\right),
\end{equation}
where $L_{gr}$ is the gravitational Lagrangian and $\phi$ is a scalar field, the NEC often reduces to \cite{Alkac:2018whk}
\begin{equation}
    (-T_t^t + T^r_r) = \frac{1}{2} \phi'^2 \ge 0. \label{con}
\end{equation}
Since the right-hand side (r.h.s) is always positive definite, therefore, the candidate c-function\footnote{The authors in \cite{Alkac:2018whk} compute the c-function using the null energy condition for the Einstein gravity in 3D with negative cosmological constant and obtain the correct central charge for the dual boundary theory $CFT_2$. Following a similar approach, we will compute the central charge associated with the sDG theory. However, the explicit computation of the c-function for the DSSYK model has not yet been accomplished.} can be defined by an integral of the r.h.s of (\ref{con}). This ensures the monotonicity of the c-function as we elaborate below with a specific example of sDG.\par
To begin with, we compute the stress-energy tensor for the right sector of the full theory (\ref{lact1}) and obtain the corresponding flux density \cite{Penrose:1964wq}, which is further used to construct the right c-function. Finally, one can perform a similar calculation for the left sector of (\ref{lact1}) and obtain the associated left c-function.


The stress energy tensor for the right sector of (\ref{lact1}) is given by
\begin{align}
 T^{r}_{\mu\nu} = \frac{2}{\sqrt{-g}}\frac{\delta S_r(Z_1)}{\delta \hat{g}^{\mu\nu}} = \frac{1}{2\pi}(\partial_\mu Z_1 \partial_\nu Z_1 - \frac{1}{2}\hat{g}_{\mu\nu} \hat{g}^{\rho\sigma}\partial_\rho Z_1\partial_\sigma Z_1) - \mu_1 \hat{g}_{\mu\nu}e^{2b_1Z_1}.
 \end{align}
Notice that in order to compute the stress-energy tensor, we first replace the flat space-time metric ($\hat{g}_{\mu\nu}$) with a dynamic metric and then take the variation of the action. After performing the variation, we substitute the dynamic metric back with the flat space-time metric. 
 
Next, we choose a null vector $k^\mu = (1,1)$ that satisfies the condition $k_\mu k^\mu = 0$ and compute the following quantity
\begin{align}
    T_r = T^{r}_{\mu\nu} k^\mu k^\nu =\frac{1}{4\pi}k^\mu k^\nu \partial_\mu Z_1 \partial_\nu Z_1 =\frac{1}{4\pi} Z_1'(u)^2.\label{t}
\end{align}
Here, $T_r$ is proportional to the flux density (radial kinetic energy density) \cite{Alkac:2018whk} in the null direction $k^\mu$.

Using (\ref{t}), we now define the total flux ($E_r$) for the right sector 
\begin{align}\label{er}
    E_r = \int d^2x  T_r.
\end{align}

The corresponding flux density turns out to be
\begin{align}
  \epsilon_r(u)= \frac{dE_r}{du}  = \frac{t_0}{4\pi} |(Z_1')^2|, \label{fluxdef} 
\end{align}
where $t_0$ is the total time interval that comes due to the integration over the time coordinate. \par 
Given the entity as in (\ref{fluxdef}), one can easily relate the NEC approach to that with the superpotential approach by proposing
\begin{align}
    D(u) = \frac{4\pi}{t_0}|e^{(Z_1 + iZ_2)/4} \epsilon_r(u)|,
\end{align}
which is precisely the function (\ref{ed}) obtained previously. We can perform similar analysis for the left sector. This defines the c-functions for the left and the right sectors as defined in (\ref{cm}). Adding the two c-functions would then give us the total c-function which is the same as (\ref{cfw}). However, defining the total c-function from the sDG action (\ref{lact1}) must be done carefully. The total stress-energy tensor is given by
\begin{align}
    T_{\mu\nu} =&\hspace{1mm} \frac{1}{2\pi}(\partial_\mu Z_1 \partial_\nu Z_1 - \frac{1}{2}\hat{g}_{\mu\nu} \hat{g}^{\rho\sigma}\partial_\rho Z_1\partial_\sigma Z_1) - \mu_1 \hat{g}_{\mu\nu}e^{2b_1Z_1} +  \frac{1}{2\pi}(\partial_\mu Z_2 \partial_\nu Z_2 \nonumber\\&\hspace{1mm} - \frac{1}{2}\hat{g}_{\mu\nu} \hat{g}^{\rho\sigma}\partial_\rho Z_2\partial_\sigma Z_2) - \mu_2 \hat{g}_{\mu\nu}e^{2b_2Z_2}. \label{totalt}
\end{align}
\par
Contracting (\ref{totalt}) with a null like vector produces,
\begin{align}
    T = \frac{1}{4\pi}(Z_1'(u)^2 + Z_2'(u)^2) = T_r + T_l .\label{totalT}
\end{align}
\par
Earlier, we have defined the c-function for each of the sectors (say for the right sector)  by first defining a density function
\begin{align}
    D(u) \propto |T_r|
\end{align}
and thereby integrating it.
\par
However, we cannot obtain the total density function ($D\text{(right sector)} + D\text{(left sector)}$) from $T$ (\ref{totalT}) following the same procedure since
\begin{align}
    |T_r| + |T_l| \neq |T|.
\end{align}
Moreover, taking the modulus of $T$ produces a function that vanishes in the UV. In other words, one cannot obtain  the correct c-function by integrating $|T|$. Therefore, the correct way of defining the total c-function is the previous method of treating  both the sectors separately and then taking the sum.\par

\subsection{Other approaches of constructing the c-function : a comparative study}
Here, we summarize other approaches to c-function and their shortcomings for 2D sine dilaton gravity. In literature, various methods have been proposed to construct the holographic c-function. Notable examples include the covariant approach \cite{Sahakian:1999bd} and the approach based on top down construction \cite{Jokela:2025cyz}-\cite{Chatzis:2025dnu}. In the following we discuss them briefly.

The authors in \cite{Sahakian:1999bd} compute the holographic c-function using a covariant approach in $D$-dimensional space-time. This method relies on geometrical quantities and is independent of the choice of coordinates. In particular, they picked a null congruence\footnote{The null congruence condition is defined as $n^an_a=0$ and $T_{ab}n^an^b>0$, where $T_{ab}$ is the energy-momentum tensor.}, defined as the family of light rays in a $D-2$ dimensional surface. Let $n^a$ be the tangent vector along these geodesics. Next, they define a scalar quantity ($\xi$) using the tangent vector ($n^a$) and an induced metric ($h^{ab}$) on $D-2$ hypersurface as
\begin{align}
    \xi=h^{ab}\nabla_an_b.\label{xi}
\end{align}
These geodesics will converge when $\xi\leq0$ \cite{Bousso:1999xy}.

Finally using (\ref{xi}), they propose the following c-function ($c(\tau)$)
\begin{align}
    c(\tau)=\frac{c_0}{G_D\int\sqrt{h}\xi^{D-2}\Big|_{\tau}}\label{ccf},
\end{align}
where $\tau$ is the proper time, $G_D$ is the gravitational constant in $D$ dimensions and $c_0$ is a numerical constant. Using the Raychaudhuri's equation, they further show that the c-function (\ref{ccf}) is a monotonic decreasing function of $\tau$, provided $\xi\leq0$ and the null congruence condition is satisfied.

Notice that the above approach fails for $D=2$, which forces us to search for alternatives in case of 2D gravity models. One such approach has been proposed recently in \cite{Suh:2020qnl}, which serves as the basis of our present calculation. 

Alternatively, there exists top down construction \cite{Jokela:2025cyz}-\cite{Chatzis:2025dnu} for computing the holographic c-function. Such methods explicitly rely on the geometry of the internal manifold. To construct a c-function following this approach, as a first step, one has to uplift sine-dilaton gravity in 10D, which is beyond the scope of the present paper and is clearly an interesting project to be pursued in the future.

Finally, it is worth noting that in higher-dimensional holographic RG flows, the entropic c-functions can be constructed from the entanglement entropy of spatial subregions, and their monotonicity is closely related to the NEC as discussed by the authors in \cite{Myers:2012ed}. However, in the context of $AdS_2$/$CFT_1$ holography, there is no canonical spatial interval on the boundary, and entanglement entropy is instead described by the generalized entropy \cite{Almheiri:2019psf}-\cite{Almheiri:2019qdq}. Despite the extensive literature on generalized entropy and entanglement entropy in JT gravity, a precise entropic c-function analogous to those in higher-dimensional holography has not yet been established. Nevertheless, we anticipate a close relationship between the aforementioned c-function and the generalized entropy. Establishing a precise connection would require a thorough investigation of entanglement entropy in sDG, which we leave for future work.



 \section{Conclusion and future direction}

We conclude our analysis with an emphasis on the fact that the sDG reduces to the JT gravity in the deep IR limit. The authors in \cite{Blommaert:2024whf} have discussed that at low energies, the DSSYK model reduces to the SYK theory, which occurs when we take the limit as $q\rightarrow 1$. However, in the bulk, the corresponding reduction of the sDG to JT gravity can also be described in terms of a holographic RG flow, where the radial coordinate ($r$) controls the energy scale \cite{Natsuume:2014sfa}. Below, we illustrate this reduction.

To begin with, we consider the sDG action (\ref{sdgact}) given below
\begin{align}
     S_{sDG} = \frac{1}{4|\log q|}\int d^2 x \sqrt{-g}\left( R \Phi + 2 \sin\Phi\right).
\end{align}
Next, we rescale the dilaton as $\Phi \rightarrow 2\pi + 2|\log q| \Phi$ as discussed by the authors in \cite{Mahapatra:2025fpx}. We then take the limit as $|\log q| \rightarrow 0$, yielding the JT gravity action\footnote{Here, we identify $|\log q|$ with $4 \pi G_2$ \cite{Mahapatra:2025fpx}, where $G_2$ is the gravitational constant in two dimensions.}
\begin{align}\label{jtgs}
    S_{JT} = \frac{1}{16\pi G_2}\int d^2x \sqrt{-g}\Phi(R + 2).
\end{align}

Alternatively, one can obtain the JT gravity (\ref{jtgs}) at the level of equations of motion. Notice that upon solving equations of motion associated with the sDG theory (\ref{r1e})-(\ref{metric2}), we find $\Phi(r)=r$ (\ref{phisolr}). In the deep IR limit, where $r\rightarrow0$, the sine dilaton potential ($\sin[\Phi]$) simplifies to a linear potential ($\Phi$), which leads to the JT gravity setup (\ref{jtgs}). As a result, the function $F(r)$ (\ref{metricsol}) further simplifies to 
\begin{align}
     F(r) = r^2 - r_h^2,
 \end{align}
  where $r_h^2 \sim \theta^2,  \theta<<1$. In other words, the JT gravity can be obtained as the IR limit when the size of the black hole shrinks to zero (i.e., $r_h\sim \theta\sim0$). Notably, the authors in \cite{Blommaert:2024whf} further show that for small values of the black hole horizon parameter ($\th$), the entropy of the system matches exactly with that of JT gravity.  
  

The present paper strengthens the above line of arguments following an explicit computation of the holographic c-function \cite{Myers:2010tj}. The details go as follows. We first obtain the solutions of background fields in sDG in the static gauge and then express these solutions in the conformal gauge or the domain wall background \cite{Suh:2020qnl} using a suitable coordinate transformation. The analysis of the holographic c-function becomes quite elegant in the domain wall background.

Next, we show the equivalence between the sDG theory and two copies of LCFT. We compute the central charge for both sectors of the LCFT and argue that the total central charge $c_{sDG}$ is the sum of the central charges pertaining to the right and the left LCFTs. 

Next, we establish the above arguments through an explicit construction of the holographic c-function using two different approaches, i.e., the superpotential formalism \cite{Suh:2020qnl} and the null energy condition approach \cite{Alkac:2018whk}. We show that both approaches coincide and yield a unique c-function. Finally, we demonstrate that in the UV limit, the c-function produces the central charge of the sDG. On the other hand, in the deep IR limit, where the sDG theory reduces to the pure JT gravity \cite{Blommaert:2024ydx}, the c-function vanishes identically \cite{Mahapatra:2025fpx}. To summarize, we have an interpolating geometry that interpolates between sDG in the UV and JT gravity in the IR and is smoothly connected by a holographic flow central charge (\ref{cfw}).

Below, we outline a few interesting projects that can be pursued in the future.

$\bullet$ Recently, the authors in \cite{Mahapatra:2025fpx} investigated the holographic properties of the sDG theory when coupled with a U(1) gauge field. In particular, they compute the central charge for the theory in both the UV and IR limits. It would be an interesting project to repeat the above calculations in the presence of gauge interactions and compute the associated holographic c-function.

$\bullet$ The authors in \cite{Aguilar-Gutierrez:2024oea} explored the $T\bar{T}$ deformation in the context of the sDG model. It must be an interesting project to address how such deformations impact the RG flow or the c-function.

    \section*{Acknowledgments}
    PM and DR are indebted to the authorities of Indian Institute of Technology, Roorkee for their unconditional support towards researches in basic sciences. HR would like to thank the authorities of Saha Institute of Nuclear Physics, Kolkata, for their support. The authors would like to thank Prithvi Narayan for his useful discussions. DR also acknowledges the Mathematical Research Impact Centric Support (MATRICS) grant (MTR/2023/000005) received from ANRF, India.
\appendix
\section{Asymptotics of the integral function}\label{app}
In this Section, we examine the asymptotic behaviour of the c-function (\ref{cfw}) associated with the Liouville CFT. In particular, we study this function in both the UV ($u\rightarrow0$) and the IR ($u\rightarrow\infty$) regimes.

The c-function function of the LCFT (\ref{cfw}) is given below
\begin{align}
    \hat{c}(u)=\frac{c}{N}\int_u^\infty D(s) ds\hspace{1mm},\hspace{2mm} D(u)=\frac{4\rho_h^{3}\text{csch}^{3}(\rho_h u)}
{\big(1+\rho_h^{2}\text{csch}^{2}(\rho_h u)\big)^{3/2}}.\label{if}
\end{align}

To begin with, we first compute the integration (\ref{if}) in the UV limit, where $u\rightarrow0$. In this limit, $D(u)$ simplifies to
\begin{align}
    D(u)\Big|_{u\rightarrow0}= 4-6 u^2+O\left(u^4\right).\label{duv}
\end{align}








Next, we integrate $D(u)$ term by term as follows
\begin{align}\label{iu1}
    \hat{c}(u) = \frac{c}{N} \left(\int_0^\infty D(s) ds - \int_0^u D(s) ds \right).
\end{align}

Using (\ref{duv}) and (\ref{iu1}), we find
\begin{align}\label{cce}
    \hat{c}(u)\Big|_{UV}=\frac{c}{N}\left(I_0-4u+2u^3+O(u^5)\right),
\end{align}
where we denote 
\begin{align}
   I_0\equiv I(0)=\int_0^\infty D(s)ds. 
\end{align}

Notice that $I_0$ is exactly the normalization constant ($N$) defined in (\ref{Norms}). Using this identification, i.e, $I_0=N$ in the expansion (\ref{cce}), we find
 \begin{align}\label{cce1}
     \hat{c}(u)\Big|_{UV} = c \left( 1 - \frac{4 u}{N} + \frac{2u^3}{N} + O(u^5) \right).
 \end{align}
It is clear from the above expression (\ref{cce}) that in the strict UV limit, $\hat{c}(u)$ reduces to the total central charge ($c=26$) of the theory (\ref{cls}) in the UV.


Next, we compute the integral function (\ref{if}) in the deep IR limit, where $u\rightarrow\infty$. In this limit, $D(u)$ boils down to
\begin{align}
    D(u)\Big|_{u\rightarrow\infty}= 32\rho_h^3 e^{-3\rho_h u} + O(e^{-5\rho_h u}).\label{dir}
\end{align} 

After integrating the above expression (\ref{dir}), we obtain
\begin{align}\label{ifr}
   \hat{c}(u)\Big|_{IR}\approx\frac{32}{3}\rho_h^2 e^{-3\rho_h u}.
\end{align}
Notice that the c-function (\ref{ifr}) decays exponentially in the deep IR limit as expected.

 \section{NEC without gauge fixing}
 
 In this Appendix, we derive the null energy condition in general, i.e., without fixing any gauge, and show how it connects with the results (\ref{totalT}) in the domain wall background.

We start with the bulk action (\ref{sdgact}) given by
 \begin{equation}
     S = \frac{1}{4|\log q|}\int d^2x \sqrt{-g}(\Phi R + 2 \sin\Phi).\label{aact}
 \end{equation}
 
After varying the action (\ref{aact}) with respect to space-time metric ($g_{\mu\nu}$), we obtain the stress-energy tensor as follows\footnote{We absorb an overall factor ($1/4|\log q|$) in the definition of the action.}
\begin{equation}
    T_{\mu\nu} = \bigtriangledown_\mu\bigtriangledown_\nu \Phi - g_{\mu\nu} \square \Phi + g_{\mu\nu} \sin \Phi.
\end{equation}

 Next, we define the appropriate null vector $k^\mu$ and use the relation $g_{\mu\nu}k^\mu k^\nu = 0$. This results in the following the NEC condition
 \begin{equation}
  T_{\mu\nu}k^\mu k^\nu = k^\mu k^\nu \bigtriangledown_\mu\bigtriangledown_\nu \Phi \ge 0. \label{pnec}    
 \end{equation}
 Notice that in deriving the physical NEC condition, we have not chosen any particular gauge. This result is expressed in a general covariant form.
 
  Now, one can simplify the above condition (\ref{pnec}) using the domain wall gauge (\ref{dwbg}) and the equations of motion (\ref{ee1})-(\ref{ee2}), which yields
 \begin{equation}
  T_{\mu\nu}k^\mu k^\nu = ( e^{Z_1} - e^{iZ_2}) +[(Z_1')^2 + \frac{1}{4}(Z_2')^2],\label{bat1}
 \end{equation}
 where we use the field redefinitions mentioned in (\ref{defZ1})-(\ref{defZ2}).

Notice that the above expression (\ref{bat1}) contains an additional potential term with a relative minus sign. The potential term exhibits oscillatory behavior and is not necessarily a positive-definite quantity.Therefore, one can not use the above expression (\ref{bat1}) to construct a monotonic function. However, by choosing to work in the domain-wall gauge from the beginning and applying the NEC, as discussed in section 4.2, we naturally obtain only the kinetic part of the expression (\ref{bat1}). The potential-like terms disappear due to the property of the null vector $g_{\mu\nu}k^\mu k^\nu = 0$. This allows us to construct the physically meaningful holographic c-function. This method of choosing the domain wall gauge to apply the NEC has also been extensively studied by the authors in \cite{Myers:2010tj}, \cite{Alkac:2018whk}-\cite{Myers:2010xs}. Consequently, to relate this approach to the total NEC (without gauge fixing), our construction can be interpreted as the kinetic radial projection of (\ref{bat1}) in the bulk.


\begin{thebibliography}{99}

\bibitem{Zamolodchikov:1986gt}
A.~B.~Zamolodchikov,
``Irreversibility of the Flux of the Renormalization Group in a 2D Field Theory,''
JETP Lett. \textbf{43} (1986), 730-732

\bibitem{Cardy:1988cwa}
J.~L.~Cardy,
``Is There a c Theorem in Four-Dimensions?,''
Phys. Lett. B \textbf{215} (1988), 749-752
doi:10.1016/0370-2693(88)90054-8

\bibitem{Kiritsis:2019npv}
E.~Kiritsis,
``String Theory in a Nutshell: Second Edition,''
Princeton University Press, 2019,
ISBN 978-0-691-15579-1, 978-0-691-18896-6



\bibitem{Cappelli:1990yc}
A.~Cappelli, D.~Friedan and J.~I.~Latorre,
``C theorem and spectral representation,''
Nucl. Phys. B \textbf{352} (1991), 616-670
doi:10.1016/0550-3213(91)90102-4

\bibitem{Jack:1990eb}
I.~Jack and H.~Osborn,
``Analogs for the $c$ Theorem for Four-dimensional Renormalizable Field Theories,''
Nucl. Phys. B \textbf{343} (1990), 647-688
doi:10.1016/0550-3213(90)90584-Z



\bibitem{Osborn:1991gm}
H.~Osborn,
``Weyl consistency conditions and a local renormalization group equation for general renormalizable field theories,''
Nucl. Phys. B \textbf{363}, 486-526 (1991)
doi:10.1016/0550-3213(91)80030-P



\bibitem{Komargodski:2011vj}
Z.~Komargodski and A.~Schwimmer,
``On Renormalization Group Flows in Four Dimensions,''
JHEP \textbf{12} (2011), 099
doi:10.1007/JHEP12(2011)099
[arXiv:1107.3987 [hep-th]].



\bibitem{Anselmi:1997rd}
D.~Anselmi,
``Central functions and their physical implications,''
JHEP \textbf{05}, 005 (1998)
doi:10.1088/1126-6708/1998/05/005
[arXiv:hep-th/9702056 [hep-th]].


\bibitem{Alvarez:1998wr}
E.~Alvarez and C.~Gomez,
``Geometric holography, the renormalization group and the c theorem,''
Nucl. Phys. B \textbf{541} (1999), 441-460
doi:10.1016/S0550-3213(98)00752-4
[arXiv:hep-th/9807226 [hep-th]].

\bibitem{Natsuume:2014sfa}
M.~Natsuume,
``AdS/CFT Duality User Guide,''
Lect. Notes Phys. \textbf{903} (2015), pp.1-294
doi:10.1007/978-4-431-55441-7
[arXiv:1409.3575 [hep-th]].

\bibitem{Myers:2010tj}
R.~C.~Myers and A.~Sinha,
``Holographic c-theorems in arbitrary dimensions,''
JHEP \textbf{01} (2011), 125
doi:10.1007/JHEP01(2011)125
[arXiv:1011.5819 [hep-th]].

\bibitem{Myers:2012ed}
R.~C.~Myers and A.~Singh,
``Comments on Holographic Entanglement Entropy and RG Flows,''
JHEP \textbf{04} (2012), 122
doi:10.1007/JHEP04(2012)122
[arXiv:1202.2068 [hep-th]].


\bibitem{Chu:2019uoh}
C.~S.~Chu and D.~Giataganas,
``$c$-Theorem for Anisotropic RG Flows from Holographic Entanglement Entropy,''
Phys. Rev. D \textbf{101} (2020) no.4, 046007
doi:10.1103/PhysRevD.101.046007
[arXiv:1906.09620 [hep-th]].

\bibitem{Liu:2012eea}
H.~Liu and M.~Mezei,
``A Refinement of entanglement entropy and the number of degrees of freedom,''
JHEP \textbf{04} (2013), 162
doi:10.1007/JHEP04(2013)162
[arXiv:1202.2070 [hep-th]].

\bibitem{Albash:2011nq}
T.~Albash and C.~V.~Johnson,
``Holographic Entanglement Entropy and Renormalization Group Flow,''
JHEP \textbf{02} (2012), 095
doi:10.1007/JHEP02(2012)095
[arXiv:1110.1074 [hep-th]].

\bibitem{Suh:2020qnl}
M.~Suh,
``Holographic renormalization group flows in two-dimensional gravity and $AdS$ black holes,''
JHEP \textbf{07}, 209 (2020)
doi:10.1007/JHEP07(2020)209
[arXiv:2002.07194 [hep-th]].

\bibitem{deBoer:1999tgo}
J.~de Boer, E.~P.~Verlinde and H.~L.~Verlinde,
``On the holographic renormalization group,''
JHEP \textbf{08} (2000), 003
doi:10.1088/1126-6708/2000/08/003
[arXiv:hep-th/9912012 [hep-th]].

\bibitem{Skenderis:1999mm}
K.~Skenderis and P.~K.~Townsend,
``Gravitational stability and renormalization group flow,''
Phys. Lett. B \textbf{468} (1999), 46-51
doi:10.1016/S0370-2693(99)01212-5
[arXiv:hep-th/9909070 [hep-th]].

\bibitem{Freedman:1999gp}
D.~Z.~Freedman, S.~S.~Gubser, K.~Pilch and N.~P.~Warner,
``Renormalization group flows from holography supersymmetry and a c theorem,''
Adv. Theor. Math. Phys. \textbf{3} (1999), 363-417
doi:10.4310/ATMP.1999.v3.n2.a7
[arXiv:hep-th/9904017 [hep-th]].

\bibitem{Alkac:2018whk}
G.~Alka{{c}} and B.~Tekin,
``Holographic c-theorem and Born-Infeld Gravity Theories,''
Phys. Rev. D \textbf{98} (2018) no.4, 046013
doi:10.1103/PhysRevD.98.046013
[arXiv:1805.07963 [hep-th]].

\bibitem{Myers:2010xs}
R.~C.~Myers and A.~Sinha,
``Seeing a c-theorem with holography,''
Phys. Rev. D \textbf{82} (2010), 046006
doi:10.1103/PhysRevD.82.046006
[arXiv:1006.1263 [hep-th]].

\bibitem{Li:2017txk}
Y.~Z.~Li, H.~Lu and J.~B.~Wu,
``Causality and a-theorem Constraints on Ricci Polynomial and Riemann Cubic Gravities,''
Phys. Rev. D \textbf{97} (2018) no.2, 024023
doi:10.1103/PhysRevD.97.024023
[arXiv:1711.03650 [hep-th]].

\bibitem{Li:2018kqp}
Y.~Z.~Li and H.~Lu,
``$a$-theorem for Horndeski gravity at the critical point,''
Phys. Rev. D \textbf{97} (2018) no.12, 126008
doi:10.1103/PhysRevD.97.126008
[arXiv:1803.08088 [hep-th]].

\bibitem{Ghodsi:2019xrx}
A.~Ghodsi and M.~Siahvoshan,
``A Holographic Study of the $a$-theorem and RG Flow in General Quadratic Curvature Gravity,''
Eur. Phys. J. C \textbf{79} (2019) no.10, 820
doi:10.1140/epjc/s10052-019-7345-8
[arXiv:1907.03497 [hep-th]].

\bibitem{Deddo:2023pid}
E.~Deddo, J.~T.~Liu, L.~A.~Pando Zayas and R.~J.~Saskowski,
``c-functions in higher-derivative flows across dimensions,''
JHEP \textbf{08} (2023), 147
doi:10.1007/JHEP08(2023)147
[arXiv:2305.18530 [hep-th]].


\bibitem{Blommaert:2024ydx}
A.~Blommaert, T.~G.~Mertens and J.~Papalini,
``The dilaton gravity hologram of double-scaled SYK,''
[arXiv:2404.03535 [hep-th]].

\bibitem{Blommaert:2024whf}
A.~Blommaert, A.~Levine, T.~G.~Mertens, J.~Papalini and K.~Parmentier,
``An entropic puzzle in periodic dilaton gravity and DSSYK,''
[arXiv:2411.16922 [hep-th]].


\bibitem{Blommaert:2023opb}
A.~Blommaert, T.~G.~Mertens and S.~Yao,
``Dynamical actions and q-representation theory for double-scaled SYK,''
JHEP \textbf{02} (2024), 067
doi:10.1007/JHEP02(2024)067
[arXiv:2306.00941 [hep-th]].

\bibitem{Lin:2022rbf}
H.~W.~Lin,
``The bulk Hilbert space of double scaled SYK,''
JHEP \textbf{11} (2022), 060
doi:10.1007/JHEP11(2022)060
[arXiv:2208.07032 [hep-th]].

\bibitem{Berkooz:2024lgq}
M.~Berkooz and O.~Mamroud,
``A Cordial Introduction to Double Scaled SYK,''
[arXiv:2407.09396 [hep-th]].

\bibitem{Blommaert:2025avl}
A.~Blommaert, A.~Levine, T.~G.~Mertens, J.~Papalini and K.~Parmentier,
``Wormholes, branes and finite matrices in sine dilaton gravity,''
JHEP \textbf{09} (2025), 123
doi:10.1007/JHEP09(2025)123
[arXiv:2501.17091 [hep-th]].

\bibitem{Blommaert:2023wad}
A.~Blommaert, T.~G.~Mertens and S.~Yao,
``The q-Schwarzian and Liouville gravity,''
JHEP \textbf{11} (2024), 054
doi:10.1007/JHEP11(2024)054
[arXiv:2312.00871 [hep-th]].

\bibitem{Jackiw:1984je}
R.~Jackiw,
``Lower Dimensional Gravity,''
Nucl. Phys. B \textbf{252} (1985), 343-356
doi:10.1016/0550-3213(85)90448-1

\bibitem{ktt}
A.Kitaev.2015. A simple model of quantum holography, talk given at KITP strings seminar and Entanglement program, February 12, April 7, and May 27, Santa Barbara, U.S.A.

\bibitem{Teitelboim:1983ux}
C.~Teitelboim,
``Gravitation and Hamiltonian Structure in Two Space-Time Dimensions,''
Phys. Lett. B \textbf{126} (1983), 41-45
doi:10.1016/0370-2693(83)90012-6

\bibitem{Sachdev:1992fk}
S.~Sachdev and J.~Ye,
``Gapless spin fluid ground state in a random, quantum Heisenberg magnet,''
Phys. Rev. Lett. \textbf{70}, 3339 (1993)
doi:10.1103/PhysRevLett.70.3339
[arXiv:cond-mat/9212030 [cond-mat]].

\bibitem{Maldacena:2016hyu}
J.~Maldacena and D.~Stanford,
``Remarks on the Sachdev-Ye-Kitaev model,''
Phys. Rev. D \textbf{94}, no.10, 106002 (2016)
doi:10.1103/PhysRevD.94.106002
[arXiv:1604.07818 [hep-th]].


\bibitem{Verlinde:2024zrh}
H.~Verlinde and M.~Zhang,
``SYK correlators from 2D Liouville-de Sitter gravity,''
JHEP \textbf{05}, 053 (2025)
doi:10.1007/JHEP05(2025)053
[arXiv:2402.02584 [hep-th]].

\bibitem{DiFrancesco:1997nk}
P.~Di Francesco, P.~Mathieu and D.~Senechal,
``Conformal Field Theory,''
Springer-Verlag, 1997,
ISBN 978-0-387-94785-3, 978-1-4612-7475-9
doi:10.1007/978-1-4612-2256-9

\bibitem{Polyakov:1981rd}
A.~M.~Polyakov,
``Quantum Geometry of Bosonic Strings,''
Phys. Lett. B \textbf{103} (1981), 207-210
doi:10.1016/0370-2693(81)90743-7

\bibitem{Mertens:2020hbs}
T.~G.~Mertens and G.~J.~Turiaci,
``Liouville quantum gravity -- holography, JT and matrices,''
JHEP \textbf{01} (2021), 073
doi:10.1007/JHEP01(2021)073
[arXiv:2006.07072 [hep-th]].

\bibitem{Fan:2021bwt}
Y.~Fan and T.~G.~Mertens,
``From quantum groups to Liouville and dilaton quantum gravity,''
JHEP \textbf{05} (2022), 092
doi:10.1007/JHEP05(2022)092
[arXiv:2109.07770 [hep-th]].

\bibitem{Mahapatra:2025fpx}
P.~Mahapatra, H.~Rathi and D.~Roychowdhury,
``Holographic central charge for sine dilaton gravity,''
Phys. Lett. B \textbf{869} (2025), 139813
doi:10.1016/j.physletb.2025.139813
[arXiv:2502.13884 [hep-th]].

\bibitem{Rathi:2024qsy}
H.~Rathi,
``AdS{\_}2/CFT{\_}1 at finite density and holographic aspects of 2D black holes,''
[arXiv:2404.02724 [hep-th]].

\bibitem{Hartman:2008dq}
T.~Hartman and A.~Strominger,
``Central Charge for AdS(2) Quantum Gravity,''
JHEP \textbf{04} (2009), 026
doi:10.1088/1126-6708/2009/04/026
[arXiv:0803.3621 [hep-th]].

\bibitem{Castro:2008ms}
A.~Castro, D.~Grumiller, F.~Larsen and R.~McNees,
``Holographic Description of AdS(2) Black Holes,''
JHEP \textbf{11} (2008), 052
doi:10.1088/1126-6708/2008/11/052
[arXiv:0809.4264 [hep-th]].

 \bibitem{Rathi:2023vhw}
H.~Rathi and D.~Roychowdhury,
``AdS$_{2}$ holography and ModMax,''
JHEP \textbf{07} (2023), 026
doi:10.1007/JHEP07(2023)026
[arXiv:2303.14379 [hep-th]].

\bibitem{Rathi:2021aaw}
H.~Rathi and D.~Roychowdhury,
``Holographic JT gravity with quartic couplings,''
JHEP \textbf{10} (2021), 209
doi:10.1007/JHEP10(2021)209
[arXiv:2107.11632 [hep-th]].

\bibitem{Penrose:1964wq}
R.~Penrose,
``Gravitational collapse and space-time singularities,''
Phys. Rev. Lett. \textbf{14} (1965), 57-59
doi:10.1103/PhysRevLett.14.57

\bibitem{Sahakian:1999bd}
V.~Sahakian,
``Holography, a covariant c function, and the geometry of the renormalization group,''
Phys. Rev. D \textbf{62}, 126011 (2000)
doi:10.1103/PhysRevD.62.126011
[arXiv:hep-th/9910099 [hep-th]].

\bibitem{Jokela:2025cyz}
N.~Jokela, J.~Kastikainen, C.~Nunez, J.~M.~Pen{\'\i}n, H.~Ruotsalainen and J.~G.~Subils,
``On entanglement c-functions in confining gauge field theories,''
JHEP \textbf{11}, 101 (2025)
doi:10.1007/JHEP11(2025)101
[arXiv:2505.14397 [hep-th]].

\bibitem{Chatzis:2025dnu}
D.~Chatzis, M.~Hammond, G.~Itsios, C.~Nunez and D.~Zoakos,
``Universal observables, SUSY RG-flows and holography,''
JHEP \textbf{08}, 134 (2025)
doi:10.1007/JHEP08(2025)134
[arXiv:2506.10062 [hep-th]].

\bibitem{Bousso:1999xy}
R.~Bousso,
``A Covariant entropy conjecture,''
JHEP \textbf{07}, 004 (1999)
doi:10.1088/1126-6708/1999/07/004
[arXiv:hep-th/9905177 [hep-th]].

\bibitem{Almheiri:2019psf}
A.~Almheiri, N.~Engelhardt, D.~Marolf and H.~Maxfield,
``The entropy of bulk quantum fields and the entanglement wedge of an evaporating black hole,''
JHEP \textbf{12} (2019), 063
doi:10.1007/JHEP12(2019)063
[arXiv:1905.08762 [hep-th]].

\bibitem{Almheiri:2019qdq}
A.~Almheiri, T.~Hartman, J.~Maldacena, E.~Shaghoulian and A.~Tajdini,
``Replica Wormholes and the Entropy of Hawking Radiation,''
JHEP \textbf{05} (2020), 013
doi:10.1007/JHEP05(2020)013
[arXiv:1911.12333 [hep-th]].


\bibitem{Aguilar-Gutierrez:2024oea}
S.~E.~Aguilar-Gutierrez,
``$T^2$ deformations in the double-scaled SYK model: Stretched horizon thermodynamics,''
[arXiv:2410.18303 [hep-th]].

\end{thebibliography}
\end{document}